\documentclass[a4paper,12pt]{article}
\pdfoutput=1 

\usepackage{jheppub} 

\usepackage[T1]{fontenc} 
\usepackage{mathtools}

\usepackage{enumitem}

\usepackage{aas_macros}
\usepackage{amsmath,amssymb}
\usepackage{multirow}
\usepackage{cleveref}
\usepackage{siunitx}
\usepackage{enumerate}
\usepackage[dvipsnames]{xcolor}
\usepackage{float}

\usepackage[compat=1.1.0]{tikz-feynman}
\usepackage{afterpage}

\usepackage{amsthm}
\usepackage{slashed}
\usepackage{psfrag}
\usepackage{color}
\usepackage{graphicx}
\usepackage{upgreek}
\usepackage{feynmp}
\DeclareMathAlphabet{\mathpzc}{OT1}{pzc}{m}{it}

\usepackage{hyperref}
\usepackage[mathscr]{euscript}

\usepackage[toc,page]{appendix}
\usepackage{circledsteps}
\usepackage{natbib}
\setcitestyle{square,comma,numbers,sort&compress}
\usepackage{enumerate}
\usepackage{tabularx,booktabs}
\usepackage[dvipsnames]{xcolor}
\usepackage{comment}
\usepackage[caption=false]{subfig}
\usepackage[compat=1.1.0]{tikz-feynman} 

\usepackage[section]{placeins}
\usepackage{cleveref}

\allowdisplaybreaks[4]

\definecolor{darkgreen}{rgb}{0,0.5,0}

\newcommand{\beq}{\begin{eqnarray}}
\newcommand{\eeq}{\end{eqnarray}}

\newcommand{\bseq}{\begin{subequations}}
\newcommand{\eseq}{\end{subequations}}
\newcommand{\be}{\begin{equation}}
\newcommand{\ee}{\end{equation}}

\newcommand{\bra}[1]{\ensuremath{\langle #1 |}}   
\newcommand{\ket}[1]{\ensuremath{| #1 \rangle}}   

\renewcommand{\Im}{\mathop{\rm Im}\nolimits}

\newcommand{\beqa}{\begin{eqnarray}}
\newcommand{\eeqa}{\end{eqnarray}}

\def\bra#1{\langle\,#1\,\vert}
\def\ket#1{\vert\,#1\,\rangle}
\usepackage{simpler-wick}
\newcolumntype{Y}{>{\centering\arraybackslash}X}
 
\title{The equivalent Electric Dipole Moment in SMEFT}

\author{Marco Ardu,}

\author{Nicola Valori} 

\affiliation{Instituto de F\'isica Corpuscular, Universidad de Valencia and CSIC, Edificio Institutos Investigaci\'on, C/Catedr\'atico
Jos\'e Beltr\'an 2, 46980 Paterna, Spain}

\emailAdd{marco.ardu@ific.uv.es}
\emailAdd{nicola.valori@uv.es}

\abstract{The Electric Dipole Moment of the electron (eEDM) is typically investigated in experiments using paramagnetic molecules. However, the physical observable in these searches consists in a linear combination of CP-violating interactions, rather than the eEDM alone, which is commonly referred to as the equivalent EDM of the system. Assuming the presence of new CP-odd physics from heavy degrees of freedom, we parameterize its effects within the Standard Model Effective Field Theory (SMEFT) framework. We systematically compute the contributions to the full low-energy direction probed by EDM searches, focusing on leading-order effects at dimension six and one-loop level, while also discussing selected two-loop and dimension-eight contributions. We find that eEDM experiments are sensitive to a broader class of SMEFT operators than previously recognized.
}

\begin{document}
\maketitle
\flushbottom

\section{Introduction}
The Standard Model (SM) of particle physics provides a remarkably precise description of fundamental interactions, successfully accounting for most observed phenomena up to energies near the TeV scale. However, despite its successes, several key observations remain unexplained within the SM framework. These include the existence of nonzero neutrino masses, the presence of dark matter, and the observed baryon asymmetry of the universe. Addressing these issues requires physics beyond the SM (BSM), motivating both theoretical developments and experimental searches for new interactions.

In particular, the baryon–antibaryon asymmetry of the universe cannot be dynamically generated within the SM alone \cite{Gavela:1993ts, Huet:1994jb}. A key requirement for successful baryogenesis are additional sources of CP violation beyond those present in the CKM and PMNS matrices, which govern quark and lepton mixing respectively. This motivates experimental searches for CP-odd observables that could serve as direct signatures of new CP-violating interactions. Among these, Electric Dipole Moments (EDMs) of SM particles provide some of the most stringent probes of CP-violating BSM physics. Theoretically, EDMs are particularly well-suited for New Physics (NP) searches because their SM contributions are extremely suppressed. In the quark sector, CP violation is parametrized by the Jarlskog invariant \cite{Jarlskog:1985ht}, $J=\Im(V_{tb} V^*_{td} V_{cd} V^*_{cb}) \sim 3\times 10^{-5}$ \cite{ParticleDataGroup:2022pth}, which leads to EDM contributions only appearing at three-loop order or higher due to the required multiple CKM insertions and the antisymmetry of $J$ \cite{PhysRevD.89.056006}\footnote{ Long distance contributions can lead to larger values \cite{PhysRevLett.125.241802, PhysRevD.103.013001}, but are still far from the experimental sensitivites}. In the lepton sector, an analogous Jarlskog invariant can be constructed from the PMNS matrix, but no definite sign of CP violation in neutrino oscillations has been measured so far\footnote{Although there is a small preference for a non-zero Dirac phase when including SK-ATM data \cite{Esteban:2020cvm, Capozzi:2021fjo}}. Additionally, the PMNS induced EDMs also appear at high loop orders, leading to suppressed contributions.
Experiments looking for EDMs are sensitive to much larger values than those predicted by the SM, and would thus be able to unambiguously detect or constraint CP-odd new physics interactions.
\begin{table}[t]
    \centering
    \begin{tabular}{c|c}
        EDM &  Upper-Limit @ 90\% CL \\
        \hline
        $d_e$ & $4.1\times 10^{-30}\ e\cdot {\rm cm}$ \cite{Roussy_2023}\\
        $d_\mu$ & $1.8\times 10^{-19}\ e\cdot {\rm cm}$ \cite{Muong-2:2008ebm}\\
        $d_\tau$ & $\sim 10^{-17}$ \cite{Belle:2021ybo}\\
        $d_n$ & $1.8\times 10^{-26}\ e\cdot {\rm cm}$ \cite{Abel:2020pzs}\\
    \end{tabular}
    \caption{Current upper-limit on the EDMs of charged leptons and the neutron. As explained in the text, the electron EDM bounds are found assuming that the coefficient of the semi-leptonic interactions that enter in the equivalent EDM of the paramagnetic HfF+ molecule is zero. }
    \label{tab:EDMbounds}
\end{table}
The absence of EDM signals in present searches has led to stringent upper limits on the EDMs of leptons and the neutron, summarized in Table \ref{tab:EDMbounds}, with future experiments expected to further improve the sensitivities. The projection for future electron EDM searches anticipate an improvement of up to three order of magnitude \cite{Hiramoto:2022fyg, Fitch_2020, Vutha:2018tsz}. Dedicated searches for a muon EDM at PSI are also expected to run in the upcoming years, with a target sensitivity of $d_\mu\lesssim 6\times 10^{-23}\ e\cdot {\rm cm}$ \cite{Sakurai:2022tbk}. Improvement for the neutron EDM are also expected from the n2EDM experiment \cite{n2EDM:2021yah}.

Assuming that new physics lies at a high energy scale, its low-energy effects can be systematically described in a model-independent manner using the Standard Model Effective Field Theory (SMEFT). In this approach, the contributions of the heavy degrees of freedom are encoded in the Wilson coefficients of higher-dimensional operators constructed from SM fields and observables can be calculated in terms of the operator coefficients. EDMs, in particular, are related to the imaginary coefficient of the dipole operator, and are sensitive to a wide variety of CP-odd operators that can contribute to it. A number of SMEFT analyses of EDMs exist in the literature, including complete studies of one-loop effects at dimension six \cite{ Aebischer_2021, Kley:2021yhn}, as well as selected two-loop contributions \cite{Panico:2018hal}.

Given the exceptional sensitivity of electron EDM experiments, constraints derived from these measurements often set the most stringent limits on BSM CP violation. In the EFT framework, the experimental reach of electron EDM searches corresponds to the largest UV scales among all EDM observables. The state-of-the-art of electron EDM best limit come from searches of the EDMs of paramagnetic molecules, like $\rm HfF^+$ and ThO. However, these experiments are sensitive not only to the permanent EDM of the electron, but also to other CP-odd interactions between the electron and the nuclei. They effectively constrain a combination of the electron EDM and semileptonic scalar interactions, collectively referred to as the  "equivalent EDM" of the system ($d^{\rm equiv}$). The reported upper limits on the electron EDM assume no contribution from the CP-odd scalar, an assumption that may not hold in general. If the CP-violating physics affects electron-nuclei interactions, paramagnetic EDM searches can provide direct constraints on its magnitude. For instance, it has been recently shown that paramagnetic systems can be used to bound the $\theta$ QCD term from its contribution to the semileptonic operator entering in the equivalent EDM \cite{Mulder:2025esr}. 

The goal of this work is to systematically study the full experimental direction probed by paramagnetic EDM searches within the SMEFT framework. By including all relevant renormalization group mixing effects at the one-loop level and dimension six, we aim to provide a comprehensive analysis of how electron EDM experiments constrain the space of CP-violating operators. Selected two-loop and dimension eight contribution are also briefly discussed.

The paper is organised as follows. In Section \ref{sec:equivEDM} we review the low-energy operators that enter in the definition of the equivalent EDM, and discuss the combinations probed by the different paramagnetic molecules. In Section \ref{sec:edmSMEFT} we introduce the SMEFT Lagrangian and calculate the contributions of high-energy operators onto the low-energy experimental direction to constrain the SMEFT operator coefficients. Finally, we summarize our results in Section \ref{sec:Conclusion}. 

\section{The equivalent electric dipole moment}\label{sec:equivEDM}


As discussed in the Introduction, the most sensitive searches of the electron EDM (eEDM) are conducted on paramagnetic systems, where the EDM of the unpaired electrons is inferred from the  frequency shift of paramagnetic atoms or molecules in the presence of an electric field. 
However, these energy shifts  depend not only on the eEDM, but also on other CP-odd semileptonic interactions.
At the experimental scale, the relevant interactions to which experiments are most sensitive can be parametrized according to the Lagrangian \cite{Pospelov:2005pr}:
\begin{equation} \label{Lag:lowenergy}
    \mathcal{L} = -i\frac{d_{e}}{2} \bar{e} \sigma_{\mu \nu} \gamma_{5}e F^{\mu \nu} +  \frac{G_{F}}{\sqrt{2}} C_{S} \bar{e} i \gamma_{5} e \bar{N} N,
\end{equation}
where $\sigma_{\mu \nu} = \frac{i}{2} \left[\gamma_{\mu},\gamma_{\nu}\right]$, $N = (p,n)$ is the nucleon isospin doublet, $G_F$ is the Fermi constant and $C_{S}$ is a dimensionless Wilson coefficient. It is possible to define an equivalent eEDM 
 ($d_{e}^{\mathrm{equiv.}}$) in terms of operators in (\ref{Lag:lowenergy}) that correspond to the observable measured in   paramagnetic EDM searches. Currently, the best experimental sensitivity is obtained using $\mathrm{HfF^{+}}$ molecular ions, where the equivalent EDM reads\footnote{We are neglecting the isospin breaking effects that are proportional to the coefficient of the semileptonic operator $\bar{e} i \gamma_{5} e \bar{N}\tau^3 N$, with $\tau^3$ being the diagonal Pauli matrix, because they are suppressed by the small  $Z-N$ difference in the molecule, where $Z$ is the number of protons and $N$ the number of neutrons} \cite{Chupp:2017rkp}
 \begin{align} \label{Eq:equiv.}
    &d_{\rm HfF^+}^{\mathrm{equiv}}= d_{e} + C_{S} \times 0.9 \times 10^{-20}\ e\cdot {\rm cm}.
\end{align}
and is constrained to be 
 \cite{Roussy_2023}:
 \begin{equation}\label{edm_limit}
    \vert d_{\rm HfF^+}^{\mathrm{equiv}}\vert < 4.1 \times 10^{-30}\ e\cdot {\rm cm}.
\end{equation}
Within the SM, $d_{e}^{\rm equiv}$ is predicted to be $\sim 10^{-35}\, e \cdot {\rm cm}$ \cite{PhysRevLett.129.231801}, which is still far from the reach of future eEDM experiments \cite{Hiramoto:2022fyg,Vutha:2018tsz,Fitch_2020}. It is then possible to constrain CP-violating operators from the $d_{e}^{\mathrm{equiv}}$ upper-limit.

A single experiment cannot distinguish the contributions of the two operators in Eq.~(\ref{Lag:lowenergy}), making it impossible to determine whether a potential signal arises from $d_e$ or $C_S$ or to impose a model-independent bound on each coefficient. When considering one-operator-at-a-time and assuming the UV matching
\begin{equation}
    d_e\sim \frac{v}{16\pi^2 \Lambda^2}\qquad C_S\sim \frac{1}{\Lambda ^2 G_F},
\end{equation}
where $v\sim 174$ GeV is the electroweak Vacuum Expectation Value (VEV), 
the eEDM still remains the most sensitive observable to heavy new physics, given that it can probe energy scales $\Lambda\lesssim$ $10^{7}$ TeV which are $10 \sim 10^{2}$ times larger than those probed by the semileptonic contribution.
However, the dipole coefficient could be suppressed by small couplings or loop factors, potentially allowing the scalar contribution to compete or dominate in $d^{\rm equiv}$. 

Since the low-energy EFT space spanned by the eEDM and the semileptonic scalar operator is two-dimensional, a pair of experiments using different molecules would, in principle, be sufficient to resolve the degeneracy. This holds true provided that the experimental sensitivities define sufficiently orthogonal directions in the coefficient plane; otherwise, both experiments would probe nearly the same linear combination of operators, leaving the distinction ambiguous.

Searches for a permanent EDM in thorium monoxide ThO are sensitive to the equivalent EDM combination
\begin{equation}
        d^{\rm equiv}_{\rm ThO}=d_e+C_S\times 1.5\times 10^{-20}\ e\cdot {\rm cm}
\end{equation}
which is currently bounded to be $ < 1.1 \times 10^{-29} \ \mathrm{e} \cdot \mathrm{cm}$ \cite{ACME:2018yjb}. The ${\rm HfF^+}$ direction
can be decomposed into a projection onto $d^{\rm equiv}_{\rm ThO}$ and the perpendicular direction as
\begin{equation}
    d_{\rm HfF+}\sim 0.75 (\cos\phi\ d_{\rm Tho}+\sin\phi\ d_{\rm Tho\perp}) \label{eq:directionperp}
\end{equation}
where $\sin\phi\simeq 0.25$ and the orthogonal combination is
\begin{equation}
    d_{\rm Tho\perp}=1.5\ d_e-C_S\times 10^{-20}\ e\cdot {\rm cm}.
\end{equation}
From Eq.~(\ref{eq:directionperp}), in case of successful detection of $d^{\rm equiv}$, the two directions can be disentangled if the two measurements have a relative accuracy of order $\sim \tan \phi$, rendering possible to measure $d_e$ and $C_S$ separately. A global fit of the current two independent measures leads to \cite{Roussy_2023}:
\begin{equation}
    \vert d_{e} \vert < 2.1 \times 10^{-29}\ e \cdot {\rm cm} \quad \vert C_S\vert < 1.9 \times 10^{-9} 
\end{equation}
at $90 \% \; \mathrm{C.L.}$, which is roughly one order of magnitude milder than the bounds inferred from the most sensitive measure assuming that one of the two coefficients is negligible. Note that the loss in sensitivity could be traced to the less precise measurement  on ThO, but even with comparable precision the loss of one-order of magnitude in the combined bound is expected from Eq.~(\ref{eq:directionperp}), because the two experimental directions  are aligned up to the small angle $\sin\phi\sim0.25$.

 \subsection{Matching at the nucleon scale}\label{ssec:matchingnucleon}
In this section, we derive the matching conditions at $\mu \sim 2$ GeV, the scale where QCD becomes non-perturbative, relating nucleon-level and quark-level operators relevant for the interactions contributing to the equivalent EDMs.  We consider operators with chiral fields because it facilites the matching with the SMEFT. 

The scalar quark operators relevant for the matching onto the CP-odd semi-leptonic operator of Eq.~(\ref{Lag:lowenergy}) are
\begin{equation}\label{Eq:semileptonic_operators}
    O_{XY}^{eq} = (\bar{e}P_{X}e) \; (\bar{q}P_{Y}q)
\end{equation}
where $P_{X,Y}$ stands for the chirality projectors with $X,Y=L,R$, and $q$ labels the SM quarks. These are added to the Lagrangian as
\begin{equation}
    \mathcal{L}_q= \sum_{q}\sum_{X,Y=L,R} \frac{C^{eq}_{XY}}{\Lambda^2}O_{XY}^{eq}\
\end{equation}
where we have divided for the appropriate power of the energy scale $\Lambda$ to have dimensionless coefficients. Starting from the chiral operators, the interaction featuring a leptonic pseudoscalar current and a quark scalar current corresponds to the combination 
\begin{equation}\label{Eq:semileptonic}
    \frac{i}{2} \,\mathrm{Im}\left[C_{RL}^{eq}+C_{RR}^{eq}\right] (\bar{e} \gamma_{5} e )(\bar{q}q) \equiv C_{s}^{q} (\bar{e} i\gamma_{5} e )(\bar{q}q),
\end{equation}
\begin{table}[t]
    \centering
    \begin{tabular}{c|c|c|c}
    
       & $q=u$  & $q=d$ & $q=s$ \\
       \hline
       $G^{p,q}_S$  & 9 & 8.2 & 0.42\\
       $G^{n,q}_S$  & 8.1 & 9 & 0.42\\
    \end{tabular}
    \caption{Values of the scalar matching coefficients from \cite{Davidson:2020hkf}. The values for $u,d$ result from $\chi$PT calculations \cite{Hoferichter:2015dsa, Alarcon:2011zs}, while for $q=s$ lattice results are also used \cite{Junnarkar:2013ac}}
    \label{tab:scalarmatching}
\end{table}
since the Hermicity of the Lagrangian imposes that $C^*_{XY}=C_{\overline{X}\overline{Y}}$, where $\overline{L}=R$ and $\overline{R}=L$.

\begin{figure}
    \centering
    \includegraphics[width=0.5\linewidth]{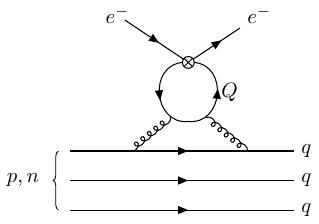}
    \caption{Contribution to the semi-leptonic interaction $C_S$ from closing a heavy quark loop in a CP-odd scalar contact interaction, and gluons are exchanged with the nucleons.}
    \label{fig:scalargg}
\end{figure}

Operators with valence quarks ($q=u,d,s$) match directly onto the nucleon-level operators because the scalar light-quark currents have non-zero matrix elements with the nucleon state, $\bra{N} \bar{q}q\ket{N} \sim G_{S}^{N,q}  \bra{N} \bar{N} N\ket{N}$, where the QCD form factors $G_{S}^{N,q}$ are given in Tab.~\ref{tab:scalarmatching}. Scalar operators with heavy quarks ($Q=c,b,t$) instead match at the one-loop level with the dimension-seven gluon operator $O_{eG} = (\bar{e} i\gamma_{5}e)(G_{\mu \nu}^{a} G^{\mu \nu}_{a})$ \cite{Davidson:2020ord}
\begin{equation}
    C_{eG}=-\frac{\alpha_{s}(m_{Q})}{12 \pi^{2} m_{Q}} C_{s}^{Q}(m_{Q}),
\end{equation}
and the gluon bilinear in turn matches onto the nucleon current at the confinement scale $\mu\sim 2$ GeV \cite{Davidson:2020ord,Cirigliano:2009bz,Shifman:1978zn}:\footnote{These contributions are depicted in the diagram of Fig.~\ref{fig:scalargg}. Note that the scalar contact interactions represented in the diagram could also result from CP-odd Higgs couplings stemming from modified Yukawa interactions}
\begin{align}\label{eq:gluonmatching} 
    &\bra{N} G_{\mu \nu}^{a} G^{\mu \nu}_{a}\ket{N} = - \frac{8 \pi m_{N}}{9 \alpha_{s}(\mu)} \bra{N} \bar{N} N\ket{N}.
\end{align}
Taking into account the QCD running of the gluon operator coefficient $C_{eG}$ (which has the same anomalous dimension as $\alpha_{s}$), we can write down a general expression for the matching onto the CP-odd scalar with nucleons:

\begin{equation} \label{Eq:semileptonic_coefficients}
    \frac{\Lambda^2 G_F}{\sqrt{2}} C_{S}(\mu)=\sum_{q=u,d,s}G_{s}^{N,q}C_{s}^{q}(\mu) + \sum_{Q=c,b,t} \frac{2 m_{N}}{27 m_{Q}}C^Q_{s}(m_{Q})
\end{equation}

\section{Equivalent EDMs in SMEFT}\label{sec:edmSMEFT}
\begin{figure}[t]
    \centering
    \includegraphics[width=1\linewidth]{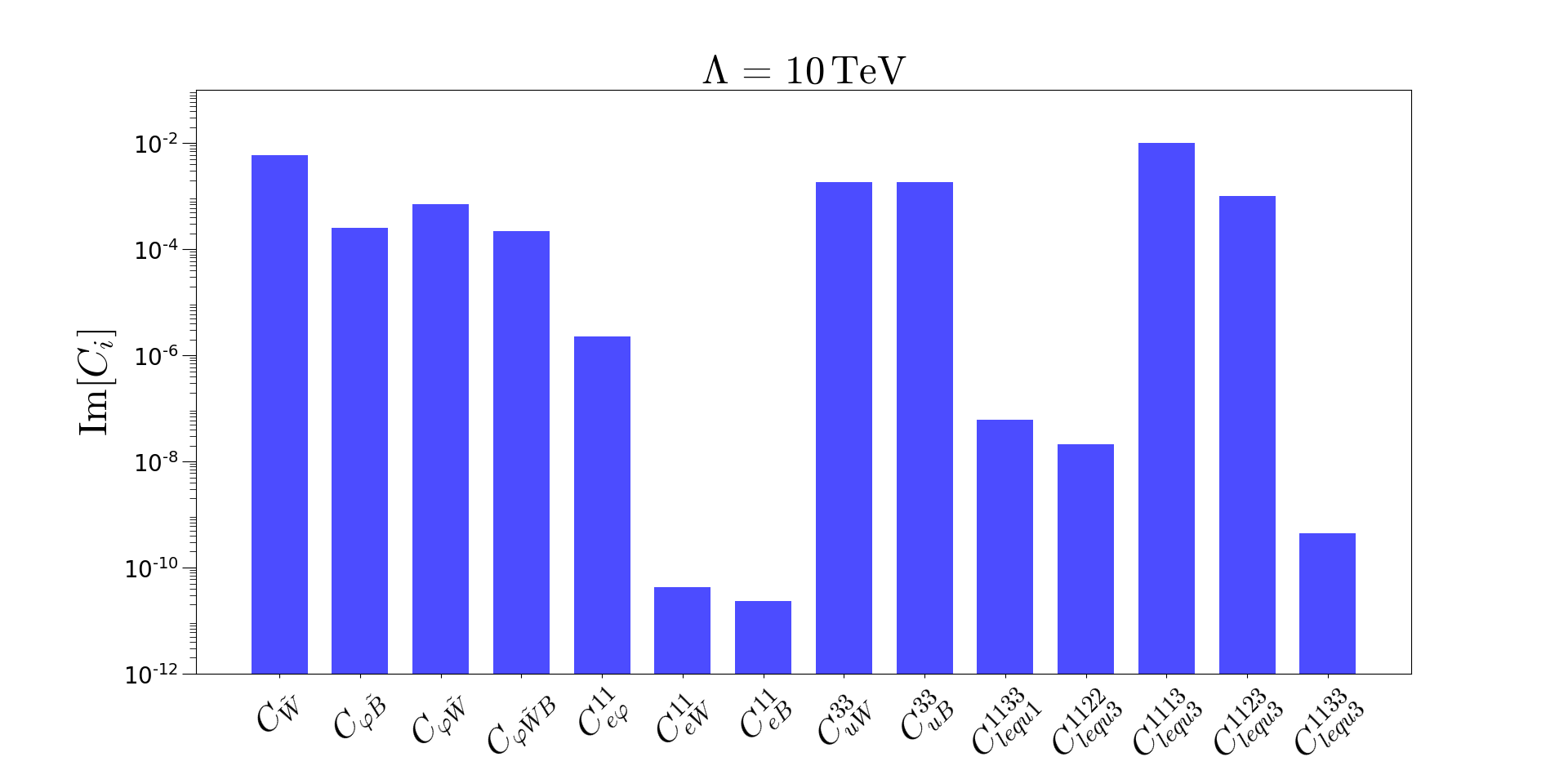}
    \caption{Sensitivities on the imaginary part of the SMEFT Wilson coefficients for the operators that primarly contribute to $d_e$, from the  EDM limit in Eq.~(\ref{edm_limit}), assuming $\Lambda$ = 10 TeV as the initial scale for the running. The first four columns refer to the real part of the correspondent Wilson coefficients, due to the fact that the violation of CP stems from the operators themselves and not from an imaginary coupling.}
    \label{fig:bound_edm}
\end{figure}

Assuming the presence of heavy physics above the electroweak scale, we can parametrize the low-energy effects with the SMEFT Lagrangian:

\begin{equation}\label{Eq:SMEFT}
    \mathcal{L}_{\rm SMEFT} = \mathcal{L}_{\rm SM} + \sum_{d} \sum_{i,\xi} \frac{C_{i}^{\xi}}{\Lambda^{d-4}} O_{i}^{\xi},
\end{equation}
where $i$ runs over all the operators of a given dimension $d$ and $\xi$ represents the flavour indices. We define the SM fields and interactions in the notation of \cite{Aebischer:2018bkb}. Here, we neglect the contributions of operator with dimension higher than six, which we take in the on-shell Warsaw basis \cite{Buchmuller:1985jz, Grzadkowski:2010es}. Dimension eight contributions are discussed in Section \ref{ssec:dimensioneight}.
We define left-handed quark doublets in the up-basis, where the up Yukawa matrix is diagonal and the quark doublets are defined as
\begin{equation} \label{Eq:ckm}
    q_{L} = 
\begin{pmatrix}
u_{L} \\
d_{L}^{\prime}
\end{pmatrix}; \quad \quad d_{L,i}^{\prime}=V_{\rm CKM}^{ij} d_{L,j},
\end{equation}
where primed fields are the down-type quarks in the weak basis, while the unprimed are the corresponding fields in the mass eigenstate basis. Flavour change involving the up-sector is possible in matching at the electroweak scale because the operator $\mathcal{O}_{uH}$ contributes to the quark masses when the Higgs acquires a VEV. Note that quark flavour mixing already occurs at the renormalizable level, because CKM causes the matrix $Y_d$ and $Y_u$ to rotate with the renormalization scale, and also contribute to off-diagonal wave function corrections. Since we work in the up-basis, the flavour changing effects are controlled by matrix elements of $Y^\dagger_dY_d$, which due to the small down-type Yukawas leads to relatively small rotation angles. For this reason, although we rotate back to the up-basis at lower scales, we do not find interesting flavour changing effects from the dimension four Renormalization Group Equations (RGEs)\footnote{With the expection of operators mixing with the top tensor $C^{eett}_{l e qu(3)}$, which we include in Fig.~\ref{fig:bound_edm}}.

We also consider singlet and doublet lepton fields in the charged lepton mass eigenstate basis, so the lepton Yukawas can have off-diagonal entries in the presence of the operator $\mathcal{O}_{eH}$. However, we consider only lepton flavour conserving operators in our analysis.

Contact interactions dressed with SM loops generally lead to divergences, which, after renormalization, cause the operator coefficients in Eq.~(\ref{Eq:SMEFT}) to evolve with the renormalization scale according to their RGEs. The solution to the dimension six RGEs, which are fully known at the one-loop level \cite{Alonso:2013hga, Jenkins:2013zja, Jenkins:2013wua}, takes the following form
\begin{equation}
    \vec{C}(\mu_f)=\vec{C}(\mu_i)U(\mu_f,\mu_i)
\end{equation}
having aligned the operator coefficients in a row vector $\vec{C}$ and where $U(\mu_f,\mu_i)$ is an evolution matrix that is related to the anomalous dimension of the coefficients. 
Hence, operators at the high scale will in general mix with each other at lower energy, allowing to constrain operators that are otherwise difficult to probe directly.

Starting from a single SMEFT operator at a time at the scale $\mu \sim \Lambda$, we run down to the electroweak scale, where the heavy SM  particles ($t$ quark and $H$, $W$ and $Z$ bosons) are integrated out. At this stage, the operators generated in the SMEFT RGEs evolution are matched onto effective interactions that are invariant under $SU(3)_c \times U(1)_{\rm em}$. Then, we continue running the low-energy operators down to the relevant energy scales, and we subsequently match onto the EDM experimental direction of Eq.~(\ref{Eq:equiv.}). The matching onto $C_S$ is given in Eq.~(\ref{Eq:semileptonic_coefficients}), where semileptonic scalar operators involving light quarks are evolved down to the confinement scale $\mu \sim 2$ GeV, while the coefficients of operators involving heavy quarks cease running at their respective mass thresholds. Note that in using Eq.~(\ref{Eq:semileptonic_coefficients}) we are not including the running of $C_S$ from 2 GeV to the electron mass $\mu\sim m_e$, which is the scale of the EDM experiments. We comment on the impact of some sub-GeV effects later in the text.

Below the electroweak scale, we parametrise the dipole operator with the electrons as:
\begin{equation}\label{Eq:dipolo}
     \mathcal{L}_{dip} = C_{e \gamma} \,\bar{e} \sigma_{\mu \nu} P_{R}e F^{\mu \nu} + h.c.,
\end{equation}
which directly matches onto the eEDM of Eq.~(\ref{Lag:lowenergy}) through the relation
\begin{equation}\label{eq:dipolechiral}
    d_{e} ={-}2 \,\mathrm{Im}[C_{e \gamma}(m_e)],
\end{equation}
where the operator coefficient $C_{e\gamma}(m_e)$ is at the electron mass scale.

The full one-loop running, as well as the tree level matching between the SMEFT and the Low Energy EFT, is performed using the public package \emph{wilson} \cite{Aebischer_2018}.

We take $\Lambda = 10$ TeV as the initial energy scale for the running and normalization of the operators. The one-loop RGE evolution is computed numerically, so it includes effects beyond the leading-log approximation.

Operators for which the contribution to $d_e$ dominates over the semileptonic term $C_S $ are shown in Fig.~(\ref{fig:bound_edm}). For these operators, considering $d^{\rm equiv}$ or $d_e$ as the experimental observable does not change the results. The operators $O_{\varphi \tilde{V}(V)}$ and $C_{l equ(3)}$ are the only ones that mix directly with the dipole at one-loop. In contrast, operators such as $O_{l equ(1)}$, $O_{\tilde{W}}$ and $O_{fV}$, with $f \neq e$, contribute to the dipole operators via second-order effects in the one-loop RGEs. We obtain results compatible with \cite{Kley:2021yhn}, although we relaxed the assumption of minimal flavour violation so we do not define chirality-flipping operators with a built-in Yukawa suppression. 
In general, we do not account for finite one-loop mixing effects\footnote{With the only exception of $O_{\tilde{W}}$, where the one-loop finite mixing has been computed in Dimensional Regularization (DR) and it is found to be comparable to the second order contribution in the one-loop RGE. While the finite contribution depends on the regularization scheme, we stress out that the only scheme consistent with an EFT approach is DR \cite{PhysRevD.43.2223,PhysRevD.89.076004}.}, as they are known to depend on the renormalization scheme, the choice of evanescent operators \cite{Herrlich:1994kh} and the treatment of $\gamma_5$ \cite{Ciuchini:1993fk}. This dependence introduces ambiguities in the UV matching and would require the inclusion of the two-loop RGEs to obtain scheme-independent observables. Selected two-loop effects have been computed in \cite{Panico:2018hal}, including the mixing of the vector operator $C^{e\tau\tau e}_{l e}$ and the scalar operator $C^{eebb}_{l e dq}$ with the dipole via insertions of the $\tau$ and $b$ Yukawa couplings. The operators \( C^{e\tau\tau e}_{l e} \) and \( C^{eebb}_{l e dq} \) also contribute to the dipole operator at higher orders in the one-loop RGEs through a chain of mixings. Specifically, they first mix with \( C^{ee tt}_{l e qu(1)} \) via a Yukawa insertion involving either the \(\tau\) lepton or \(b\) quark, which then mixes into the tensor operator with tops before ultimately contributing to the dipole. While this process is formally suppressed at three loops, it is also enhanced by a cubic logarithm, \(\log^3(\Lambda/m_t)\), whereas the direct two-loop mixing contains only a single power of \(\log(\Lambda/m_t)\). Moreover, the anomalous dimensions in the one-loop mixing chain are sizable, particularly in the tensor-to-dipole diagrams, while the two-loop mixing coefficient is comparatively suppressed. We account for the leading-log two-loop running of the scalar operator $C^{eebb}_{l e dq}$, resumming the one-loop QCD running, following the results of \cite{Panico:2018hal}. However, we find that the ratio between the two-loop contribution and the higher-order one-loop RGE-induced mixing is
\begin{equation}
    \frac{C^{\rm 1-loop,RGEs }_{e\gamma}}{C^{\rm 2-loop}_{e\gamma}}\sim 100.
\end{equation}

We also do not consider effects beyond the one-loop RGEs mixing for the vector $C^{e\tau \tau e}_{l e}$ because the result would be scheme-dependent\footnote{The authors of  \cite{Panico:2018hal} compute the two-loop mixing with the dipole of the four-lepton scalar $(\bar{l}_e e_e)(\bar{e}_\tau l_\tau)$, which is Fierz equivalent to $C^{e\tau\tau e}_{l e}$ in $d=4$ dimensions. However, in $d=4-2\epsilon$ a set of evanescent operators should be specified in order to translate the results for the vector, which may also result in one-loop finite mixing as discussed in \cite{Kley:2021yhn}.} and its contribution to $C_S$ is suppressed, so previous eEDM SMEFT analyses would apply even when considering $d^{\rm equiv}$. In contrast, $C^{eebb}_{l e dq}$ contributes significantly to the semileptonic operator featuring in $d^{\rm equiv}$, so it can be constrained directly via the matching onto the scalar.

In addition to the contributions from operators that mix with $C_{e V}$ in the RGEs, there are additional effects due to the modification of SM couplings when matching operators at the electroweak scale. Indeed, after spontaneous symmetry breaking, $C_{\psi \varphi}$ modifies the SM Yukawas, which are not anymore proportional to the fermion mass matrix. 
As a result, if $C_{\psi \varphi}$ is complex, a CP-odd interaction with the Higgs and the SM fermions survives even after rotating the fields in their mass eigenstate basis. The dominant contributions to the eEDM from modified Yukawa interactions stem from the Barr-Zee diagrams \cite{Barr:1990vd}. This is a two-loop effect where the external photon is attached to a loop of charged particles, which is then connected to the lepton line via a photon and a Higgs boson. This contribution dominates over the one-loop, which is suppressed by the small electron Yukawa, because in the Barr-Zee the Higgs couples to a heavy particle in the loop. The largest contribution arises from a top or W loop, with the CP-violating phase coming from a $C_{e \varphi}$ insertion in the electron line. 
A complex $C_{e \varphi}$ also contributes to the CP-odd scalar operators discussed in Section \ref{ssec:matchingnucleon}, when the Higgs boson is removed at the electroweak scale\footnote{Since the top is also integrated out at the same time, there is a $\propto C_{eH}$ matching contribution at one-loop to the gluon operator $(\bar{e} i\gamma_{5}e)(G_{\mu \nu}^{a} G^{\mu \nu}_{a})$, following from a diagram similar to Fig.~\ref{fig:scalargg} with a $t$ loop and the crossed dot replaced by a Higgs exchange.}.
However, since $d_{e}$ is more sensitive to heavy new physics than $C_S$, contributions to the semileptonic interaction, though sizeable, are subdominant.
\begin{figure}[t]
    \centering
    \includegraphics[width=1\linewidth]{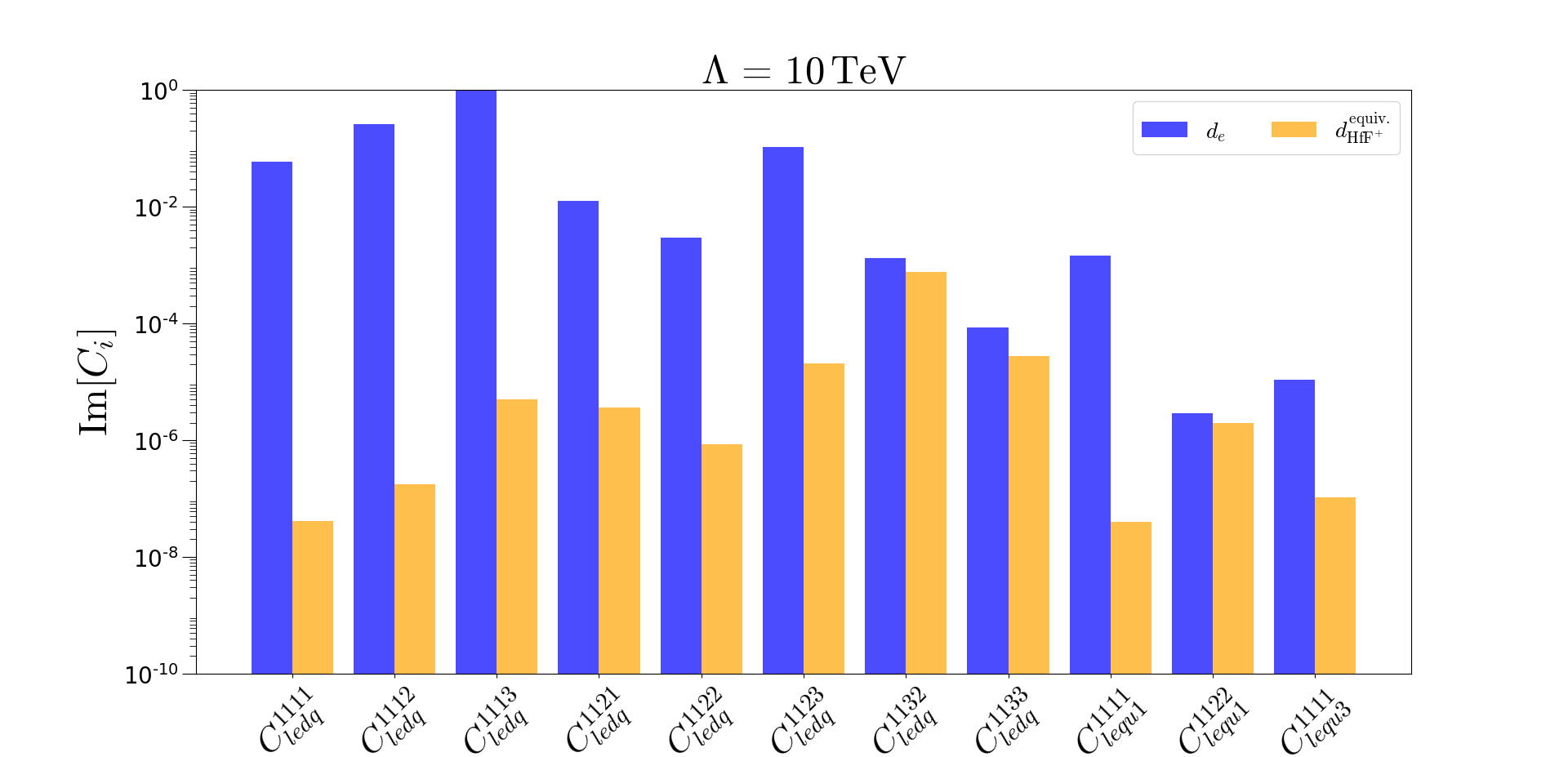}
    \caption{Comparison between the sensitivities on the imaginary part of SMEFT Wilson
coefficients from  $d_{\mathrm{HfF^{+}}}^{\mathrm{equiv.}}$ and $d_e$,
    considering the experimental limit in Eq.~(\ref{edm_limit}) and $\Lambda$ = 10 TeV as the inital scale for the running. The figure shows only the operators where the upper bound from  $d_{\mathrm{HfF^{+}}}^{\mathrm{equiv.}}$ is  stronger than from $d_e$. }
    \label{fig:bound_Cs}
\end{figure}

Fig.~(\ref{fig:bound_Cs}) compares the sensitivities on Wilson coefficients from $d_{e}$ and $d_{\mathrm{HfF^{+}}}^{\mathrm{equiv.}}$, for the operators where the contribution to $C_S$ is sizable. The semileptonic operators that are diagonal in the quark flavour match directly onto $C_S$ and the dominant RGE effect is the QCD running that rescales the operator coefficients. 
We find that the relative scaling between the different generations is in agreement with  Eq.~(\ref{Eq:semileptonic_coefficients}).
As a result, this class of operators gives the most significant contribution to $C_S$, with the only exception of $C_{l equ(1)}^{1133}$, where the corresponding matching coefficient is suppressed by $m_N/m_{t}$ and its contribution to eEDM dominates. This is due to the fact that the mixing with the dipole occurs at second-order but it is enhanced by a large anomalous dimension and features the Yukawa of the top.

Tensor operators $C_{l equ(3)}$ mix to both the eEDM and semileptonic scalar at one-loop but with different anomalous dimensions. In the former case, the one loop mixing is proportional to the Yukawa of the up-type quarks while, for the latter, the mixing occurs via electroweak gauge bosons exchange. Therefore, as a consequence of the Yukawa suppression, $d_{\mathrm{HfF^{+}}}^{\mathrm{equiv.}}$ is significantly more sensitive to operators with the up quark compared to $d_e$ alone, while for operators with tops and tensor with charms $C_S$ is negligible with respect to $d_e$. We remind that we are only including perturbative contributions above 2 GeV. However, it has been shown that sub-GeV mixing with the dipole in $\chi$PT can be significant for scalar and tensor operators involving first-generation quarks \cite{Dekens:2018pbu}. While these effects are estimated to be one to two orders of magnitude larger than the RGE-induced mixing above 2 GeV, they do not alter the conclusion that $d^{\rm equiv}$ remains the most sensitive observable, because the contributions to $C_S$ are still several orders of magnitude larger.

\subsection{Dimension eight effects}\label{ssec:dimensioneight}
Given the exceptional sensitivity of EDM experiments to NP interactions, the next to leading effects in the EFT expansion can fall within the experimental reach. These appear with dimension eight operators, which are added to the SMEFT Lagrangian as specified in Eq.~(\ref{Eq:SMEFT}), so they are suppressed by four powers of the NP scale $\Lambda$. Complete on-shell basis for dimension eight SMEFT operators can be found in \cite{Murphy:2020rsh, Li:2020gnx}, and we adopt the notation of Ref.~\cite{Murphy:2020rsh}. In this section, we only provide estimates for the contribution of selected dimension eight operators, because a complete analysis is beyond the scope of this work and would require the dimension eight RGEs, which are largerly unknown, although some partial results are available \cite{ValeSilva:2022tph, Chala:2023jyx, Chala:2023xjy, Bakshi:2024wzz, DasBakshi:2022mwk, Chala:2021pll, Davidson:2019iqh, Ardu:2022pzk, Boughezal:2024zqa}. We also disregard the contribution of derivative operators, as they are expected to be suppressed at low energies (at least at the tree-level), as well as operators that differ from a dimension-six operator only by an additional $\varphi^\dagger \varphi$ factor, since it is unlikely that a UV model would generate the dimension-eight term without also generating its dimension-six counterpart.

Tensor operators involving down-type fermions—either down-type quarks or charged leptons-- which are non-zero only if the chirality projector in the two-fermion bilinears is the same (\( P_X\otimes P_X \), with \( X=L,R \)), as well as scalar operators with the same structure, appear at leading order at dimension eight. This is because gauge-invariant interactions of this type require the insertion of two Higgs fields. The relevant operators are\footnote{The tensor operators with leptons that we consider here is Fierz equivalent to a combination of scalars $\mathcal{O}^{ijkl}_{l^2e^2\varphi^2(3)}= (l_i \varphi  e_j)(l_k \varphi  e_l)$ present in the basis of Ref.~\cite{Murphy:2020rsh}.}
\begin{align}
\mathcal{O}^{ijkl}_{l^2e^2\varphi^2(4)}= (l_i \varphi \sigma_{\alpha\beta} e_j)(l_k \varphi \sigma^{\alpha\beta} e_l)
\end{align}
\begin{align}
\mathcal{O}^{ijkl}_{l e qd\varphi^2(3)}= (l_i \varphi e_j)(q_k \varphi d_l)\qquad \mathcal{O}^{ijkl}_{l e qd\varphi^2(4)}= (l_i \varphi \sigma_{\alpha\beta} e_j)(q_k \varphi \sigma^{\alpha\beta} d_l)
\end{align}
and their respective Hermitian conjugates. The tensor operator will mix with the dimension eight dipoles with a Yukawa suppressed mixing, and contribute to the eEDM as 
\begin{equation}\label{eq:dim8tensortodipole}
    d_e\sim \frac{C_T}{16\pi^2}\frac{m_f}{\Lambda^2}\frac{v^2}{\Lambda^2}\log\left(\frac{\Lambda}{m_f}\right)
\end{equation}
where $C_T$ is the dimesion eight tensor coefficient, and $m_f$ is the mass of the respective fermion running in the loop. As in the dimension six case, the tensor operator can also mix with the associated scalar via gauge boson exchanges, which in turn will match onto four-fermion interactions with a $v^2/\Lambda^2$ suppression when the electroweak symmetry is broken\footnote{The full tree-level matching at $\mathcal{O}(1/\Lambda^4)$ between SMEFT and the low-energy EFT can be found in \cite{Hamoudou:2022tdn}.}, and contribute to the semileptonic interaction of the equivalent EDM as described in Eq.~(\ref{Eq:semileptonic_coefficients}). The tensors with quarks lighter than the strange contribute predominantly to the semileptonic operator by mixing with the scalar via gauge interactions, while the tensors with bottom quarks and $\tau$ leptons mainly contribute to the dipole as estimated in Eq.~(\ref{eq:dim8tensortodipole}). By analogy with dimension-six operators involving up-type quarks, dimension eight scalar operators with bottom quarks are expected to contribute comparably to both the semileptonic operator and the dipole, with the latter arising via an intermediate mixing with the tensor.
Another class of four-fermion operators that are not present at dimension six are scalars with up quarks having the chirality structure $P_R\otimes P_L$
\begin{equation}
    C^{ijkl}_{lequ\varphi^2(5)}=(\bar{l}_i\varphi e_j)(\bar{u}_k\tilde{\varphi}^\dagger q_l),
\end{equation}
and the respective Hermitian conjugates. These operators will match onto $C_S$ because   they give tree-level contributions to the scalars of Eq.~(\ref{Eq:semileptonic_coefficients}) with up-type quarks. Note that contrary to the dimension six case, the scalar $C^{eett}_{lequ \varphi^2(5)}$ should contribute more to $C_S$ than $d_e$, because there is no tensor with the same chirality structure to mix with as an intermediate step. \\
Finally, the operator
\begin{equation}
    \mathcal{O}^{ee}_{l e G^2\varphi}=(\bar{l}_e \varphi e)G^A_{\mu\nu} G^{A\mu\nu}
\end{equation}
matches onto the gluon operator $\mathcal{O}_{eG}$ of Section \ref{ssec:matchingnucleon} when the Higgs acquires a VEV, and contribute to the semileptonic $C_S$ at the QCD confinement scale as described by Eq.~(\ref{eq:gluonmatching}).

Additional dimension-eight effects include flavour-changing operators in the up-quark and lepton sectors that contribute to the EDM\footnote{We do not discuss down quarks because in the up-basis the down Yukawa features CKM mixing which clearly dominates over the dimension eight effects}. This arises because the operators \(\mathcal{O}_{e\varphi}\) and \(\mathcal{O}_{u\varphi}\) modify fermion masses when electroweak symmetry is broken, leading to non-diagonal Yukawa matrices in the mass eigenstate basis. Consequently, if the SMEFT is initially defined in a basis where up-type quark or charged-lepton Yukawas are diagonal, matching onto low-energy interactions requires rotating the fields to their mass eigenstates, potentially introducing flavour violation. The rotation angles are determined by the size of the \(\mathcal{O}_{\psi \varphi}\) coefficient, allowing flavour-changing operators to induce contributions to the flavour-conserving interactions relevant for the equivalent EDM. For instance, if at the electroweak scale both the operator $C^{eeij}_{l e qu(1)}$ and  $C^{ji}_{u \varphi }$ are present, then
\begin{equation}
    C^{ee jj}_{l e qu(1)}(m_W)\sim C^{ee ij}_{l e qu(1)}C^{ji}_{u\varphi}\frac{v^3}{\left(\max(m_i,m_j)\Lambda^2\right)}. 
\end{equation}
Neglecting terms suppressed by small Yukawa couplings, the rotation angles of the quark doublets are proportional to the matrix elements $C^{ij}_{u\varphi}$ with $j>i$. In contrast, the mixing of right-handed up-type quarks, which is not suppressed by Yukawa couplings, is governed by the size of the lower-left block, i.e the elements $C^{ji}_{u\varphi}$ with $j>i$. Note that this rotation is formally a double insertion of dimension six operators, hence a consistent treatment of these contributions may also require calculating the dimension eight RGEs proportional to the product of two dimension six operators. In the charged lepton sector a similar rotation should be perfomed in the presence of $\mathcal{O}^{ij}_{e\varphi}$, with $i\neq j$. However, we do not consider lepton flavour violating operators in this work, which we aim to explore in a future project.  
\begin{table}[t]
    \centering
    \begin{tabular}{c|c}
        Operator & $d^{\rm equiv}_{\rm HfF^+}$ sensitivity on $\Im(C)$\\
        \hline
        $C^{ee\tau\tau}_{l^2 e^2 \varphi^2(4)}$ & $\lesssim 10^{-4}$ \\
        $C^{eedd}_{l e qd \varphi^2(3)}$ & $\lesssim 10^{-5}$ \\
        $C^{eedd}_{l e qd \varphi^2(4)}$ & $\lesssim 10^{-4}$ \\
        $C^{eess}_{l e qd \varphi^2(3)}$ & $\lesssim 10^{-3}$ \\
        $C^{eess}_{l e qd \varphi^2(4)}$ & $\lesssim 10^{-2}$ \\
        $C^{eebb}_{l e qd \varphi^2(3)}$ & $\lesssim10^{-2}$ \\
        $C^{eebb}_{l e qd \varphi^2(4)}$ & $\lesssim 10^{-5}$ \\
        $C^{eeuu}_{l e qu \varphi^2(5)}$ & $\lesssim 10^{-5}$ \\
        $C^{eecc}_{l e qu \varphi^2(5)}$ & $\lesssim 10^{-3}$ \\
        $C^{eett}_{l e qu \varphi^2(5)}$ & $\lesssim 10^{-1}$ \\
        $C^{ee}_{leG^2\varphi}$ & $\lesssim 10^{-2}$\\  $C^{ee11}_{l e qu(1)} \sim C^{ee i1}_{l e qu(1)}C^{1i}_{u\varphi},\ C^{ee 1i}_{l e qu(1)}C^{i1}_{u\varphi}$ & $\lesssim  10^{-5}/y_{u_i} $\\
        $C^{ee11}_{lequ(3)}\sim C^{ee i1}_{l e qu(3)}C^{1i}_{u\varphi},\ C^{ee 1i}_{l e qu(3)}C^{i1}_{u\varphi}$ & $\lesssim  10^{-4}/y_{u_i} $\\
        $C^{ee22}_{lequ(1)}\sim C^{ee 12}_{l e qu(1)}C^{12}_{u\varphi},\ C^{ee 21}_{l e qu(1)}C^{21}_{u\varphi}$ & $\lesssim  10^{-3}/y_c $\\
        $C^{ee22}_{lequ(3)}\sim C^{ee 12}_{l e qu(3)}C^{12}_{u\varphi},\ C^{ee 21}_{l e qu(3)}C^{21}_{u\varphi}$ & $\lesssim  10^{-5}/y_{c} $\\
        $C^{ee22}_{lequ(1)}\sim C^{ee 32}_{l e qu(1)}C^{23}_{u\varphi},\ C^{ee 23}_{l e qu(1)}C^{32}_{u\varphi}$ & $\lesssim  10^{-3}/y_t $\\
        $C^{ee22}_{lequ(3)}\sim C^{ee 32}_{l e qu(3)}C^{23}_{u\varphi},\ C^{ee 23}_{l e qu(3)}C^{32}_{u\varphi}$ & $\lesssim  10^{-5}/y_t $\\
        $C^{ee33}_{lequ(1)}\sim C^{ee j3}_{l e qu(1)}C^{j3}_{u\varphi},\ C^{ee 3j}_{l e qu(1)}C^{3j}_{u\varphi}$ & $\lesssim  10^{-4}/y_t $\\
        $C^{ee33}_{lequ(3)}\sim C^{ee j3}_{l e qu(3)}C^{j3}_{u\varphi},\ C^{ee 3j}_{l e qu(3)}C^{3j}_{u\varphi}$ & $\lesssim  10^{-7}/y_t $\\
        \end{tabular}
    \caption{Estimated sensitivities of $d^{\rm equiv}_{\rm HfF^+}$ on the imaginary part of dimension eight coefficients or double insertion of dimension six operators, with operators normalized at $\Lambda=10$ TeV. The index $i$ can take values $i=2,3$ while  $j=1,2$}
    \label{tab:dim8}
\end{table}
The estimated sensivities on the dimension eight coefficients, and on dimension six flavour changing operators in the up quark sector are summarised in Table \ref{tab:dim8}.

\section{Conclusion}\label{sec:Conclusion}

The search for sources of CP violation beyond the Standard Model is essential for shedding light on the mechanism behind baryogenesis. At the present time, one of the best place to look for such effects are electron EDM experiments, which can probe NP scales up to $\sim 10^{7}$ TeV. Currently, the best upper limit on the eEDM is given by measurements on paramagnetic molecules. However, these experiments are sensitive to a linear combination of CP-violating interactions known as the equivalent EDM $d^{\rm equiv}$, which contains the eEDM and the Wilson coefficient $C_S$ of a semileptonic operator with electrons and nuclei. We review the definition of $d^{\rm equiv}$ in Section \ref{sec:equivEDM} for different paramagnetic molecules.

In this work, we study the contributions of higher dimensional SMEFT operators to the equivalent EDM of ${\rm HfF^+}$ molecular ions, which currently has the most stringent upper limit $\vert d_{\rm HfF^+}^{\mathrm{equiv}}\vert < 4.1 \times 10^{-30}\ e\cdot {\rm cm}$. To do so, we consider one SMEFT operator at a time for a fixed initial scale of $\Lambda=10$ TeV, which we run  down to the scale of the experiments to compute the low-energy contributions. The one-loop RGEs running, as well as all the matching between the different effective theories when crossing the SM particles masses,
has been performed with the public package \emph{wilson}. Selected two-loop running effects are also included.
The matching at the QCD confinement scale between quark-level and nucleon level operators relevant for the semileptonic interactions entering in $d^{\rm equiv}$ are given in Eq.~(\ref{Eq:semileptonic_coefficients}) of Section \ref{ssec:matchingnucleon}, where the effects of heavy quarks have also been included.

We divide the discussion of the results (Section \ref{sec:edmSMEFT}) in two parts.
First, we show in Fig.~(\ref{fig:bound_edm}) the sensitivity on classes of operators where the eEDM dominates over $C_S$. Our results are compatible with previous SMEFT analysis of the eEDM.
The new results of this work are summarized in Fig.~(\ref{fig:bound_Cs}), which presents the single-operator limits on the coefficients of those operators that contribute more significantly to $d_e^{\rm equiv}$ than to $d_e$ alone.
Semileptonic scalars and tensors are found to give larger contribution to $C_S$ than $d_e$, with the only exception of operators with top quarks and tensors with charms. 

Finally, in Section \ref{ssec:dimensioneight}, we discuss selected dimension eight effects  which, given the impressive sensitivities of EDM experiments, are estimated to be within the reach of the experimental searches. 

\acknowledgments
We thank Sacha Davidson for her useful comments on the manuscript.
We acknowledge financial support from the Spanish  Grant PID2023-151418NB-I00 funded by MCIU/AEI/10.13039/501100011033/ FEDER, UE  and from Generalitat Valenciana projects CIPROM/2021/054 and CIPROM/2022/66
\bibliography{references}
\newpage
\bibliographystyle{JHEP}
\end{document}